\documentclass[aps,showpacs,prd,twocolumn]{revtex4}
\usepackage{amssymb}
\usepackage{amsfonts}
\usepackage{amsmath}
\usepackage{graphicx}
\usepackage{color}
\usepackage[dvips]{epsfig}
\usepackage[dvips]{graphicx}
\usepackage{float}

\begin{document}

\title{Analytical BPS Maxwell-Higgs vortices}
\author{R. Casana$^{1}$, M. M. Ferreira Jr.$^{1}$, E. da Hora$^{1,2}$ and C.
dos Santos$^{3}$.}
\affiliation{$^{1}${Departamento de F\'{\i}sica, Universidade Federal do Maranh\~{a}o,}\\
65080-805, S\~{a}o Lu\'{\i}s, Maranh\~{a}o, Brazil.\\
$^{2}$Coordenadoria do Curso Interdisciplinar em Ci\^{e}ncia e Tecnologia,\\
Universidade Federal do Maranh\~{a}o, {65080-805}, S\~{a}o Lu\'{\i}s, Maranh%
\~{a}o, Brazil.\\
$^{3}${Centro de F\'{\i}sica e Departamento de F\'{\i}sica e Astronomia,
Faculdade de Ci\^{e}ncias da Universidade do Porto, 4169-007, Porto,
Portugal.}}

\begin{abstract}
We have established a prescription for the calculation of analytical vortex
solutions in the context of generalized Maxwell-Higgs models whose overall
dynamics is controlled by two positive functions of the scalar field, namely 
$f\left( \left\vert \phi \right\vert \right) $ and $w\left( \left\vert \phi
\right\vert \right) $. We have also determined a natural constraint between
these functions and the Higgs potential $U\left( \left\vert \phi \right\vert
\right) $, allowing the existence of axially symmetric
Bogomol'nyi-Prasad-Sommerfield (BPS) solutions possessing finite energy.
Furthermore, when the generalizing functions are chosen suitably, the
nonstandard BPS equations can be solved exactly. We have studied some
examples, comparing them with the usual Abrikosov-Nielsen-Olesen (ANO)
solution. The overall conclusion is that the analytical self-dual vortices
are well-behaved in all relevant sectors, strongly supporting the
generalized models they belong themselves. In particular, our results mimic
well-known properties of the usual (numerical) configurations, as localized
energy density, while contributing to the understanding of topological
solitons and their description by means of analytical methods.
\end{abstract}

\maketitle

\section{Introduction}

\label{Intro}

In the context of classical field theories, structures possessing
topologically nontrivial profiles are usually described as the static
solutions of the Euler-Lagrange equations in the presence of finite energy
boundary conditions \cite{n5}. In some special cases, by requiring the
minimization of the corresponding energy functional, such structures can
also be described via a set of first-order differential equations also known
as BPS\ equations \cite{n4}, which provide genuine solutions of the
Euler-Lagrange ones.

The kink is a one-dimensional topological object arising within the simplest
field model containing only a single real scalar field \cite{n0}. Regarding
higher dimensional scenarios, the vortex stands for a planar configuration
solving some radially symmetric Abelian-Higgs models \cite{n1}, whilst the
magnetic monopole is a three-dimensional spherically symmetric object
appearing in the non-Abelian-Higgs case \cite{n3}. All these solutions
possess the minimum energy possible, being stable against decaying into
their respective mesons. Moreover, it is well-known that, in order to give
rise to topological fields, the corresponding model must allow for the
spontaneous symmetry breaking mechanism, with its potential term presenting
at least two asymmetric vacua, since topological defects are known to be
formed during symmetry breaking phase transitions.

During the last years, a new kind of topologically nontrivial objects have
been intensively studied in connection with field models endowed with
noncanonical kinetic terms which change the dynamics of the overall system
in a nonusual way. It is worthwhile to point out that the motivation
regarding exotic dynamics arises in a rather natural way in the context of
the string theories. In particular, given some special constraints, field
models possessing nonusual dynamics also support minimum energy solutions;
see, for instance, Ref. \cite{o1}. Moreover, except for their nontrivial
nature, these solutions behave in the same general way as their standard
counterparts do. On the other hand, exotic kinetic terms also induce
slightly variations on the shape of the corresponding field solutions,
changing their amplitudes and/or characteristic lengths. Nonstandard field
models are defined by introducing generalizing functions on usual field
models. Detailed investigations regarding topological defects\ in the
context of these models are found in Ref. \cite{o2}. Many authors have also
studied interesting applications of these new solutions within several
different scenarios, specially involving the accelerated inflationary phase
of the universe \cite{n8} via the so-called k-essence models \cite{shutt},
strong gravitational waves \cite{sgw}, tachyon matter \cite{tm}, dark matter 
\cite{dm}, and others \cite{o}.

Besides the variations on the defect amplitudes and characteristic lengths,
these generalizing functions also provide new features for some models, as
for example, self-dual analytical solutions which certainly enriches our
understanding about integrable systems. Recently, self-dual analytical
monopoles were achieved in Ref. \cite{PLB} in the context of some
generalized Yang-Mill models \cite{pau}. These new analytical solutions,
unattainable in the absence of the modifying functions, were divided into
two different classes according to their capability of recovering (or not)
the standard 't Hooft-Polyakov result. By following the purpose of achieving
analytical solutions for topological defects, the present manuscript aims at
investigating the existence of analytical BPS vortex solutions within the
framework of the generalized Maxwell-Higgs model \cite{gv}.

The letter is organized as follows. In Sec. \ref{2}, we review some
important details regarding the generalized Maxwell-Higgs scenario. Sec. III
is devoted to describe the prescription implemented to find the analytical
BPS solutions of the generalized model. The consistence of our approach is
verified by investigating some explicit examples. In the sequel, the new
solutions are compared ANO profiles,\ allowing the identification of the
main properties acquired by them. Finally, in Sec. IV, we present our final
remarks and conclusions.

\section{The nonstandard model \label{2}}

\label{general}

We begin by reviewing the (1+2)-dimensional generalized Maxwell-Higgs model
introduced in Ref. \cite{gv}, whose Lagrangian density is%
\begin{equation}
\mathcal{L}=-\frac{f^{2}\left( \left\vert \phi \right\vert \right) }{4}%
F_{\mu \nu }F^{\mu \nu }+w\left( \left\vert \phi \right\vert \right)
\left\vert D_{\mu }\phi \right\vert ^{2}-U\left( \left\vert \phi \right\vert
\right) \text{,}  \label{1}
\end{equation}%
where $F_{\mu \nu }=\partial _{\mu }A_{\nu }-\partial _{\nu }A_{\mu }$ is
the usual field strength tensor, whilst $D_{\mu }\phi =\partial _{\mu }\phi
-ieA_{\mu }\phi $ stands for the covariant derivative. Moreover, $%
f^{2}\left( \left\vert \phi \right\vert \right) $ and $w\left( \left\vert
\phi \right\vert \right) $ are positive functions which change the dynamics
of the overall model, being called dielectric functions because they mimic
some effective electrodynamics in continuous media, as already mentioned in
the literature. So far, it was not explored the possibility of such
generalizing functions to provide exactly solvable models for vortex
configurations, being it the main motivation of this manuscript. Here, for
simplicity, all fields, coordinates and parameters are supposed to be
dimensionless.

The corresponding Euler-Lagrange equation for the gauge field is%
\begin{equation}
\partial _{\nu }\left( f^{2}F^{\nu \mu }\right) =J^{\mu },
\end{equation}%
where $J^{\mu }=iew\left( \phi \overline{D^{\mu }\phi }-\overline{\phi }%
D^{\mu }\phi \right) $\ is the generalized current vector, which is\ also
conserved ($\partial _{\mu }J^{\mu }=0$). The stationary Gauss law then reads%
\begin{equation}
\partial _{k}\left( f^{2}\partial _{k}A_{0}\right) =2e^{2}wA_{0}\left\vert
\phi \right\vert ^{2}\text{,}
\end{equation}%
being trivially verified by $A_{0}=0$, revealing that the static
configurations of the generalized model (\ref{1}) generate no electric field.

The stationary Amp\`{e}re's law can be written as (already using $A_{0}=0)$%
\begin{equation}
\epsilon _{ik}\partial _{k}\left( f^{2}B\right) =J_{i},  \label{eq:Ampere}
\end{equation}%
whilst the equation controlling the Higgs field\ is%
\begin{equation}
wD_{k}D_{k}\phi +\left( \partial _{k}w\right) D_{k}\phi -\left\vert
D_{k}\phi \right\vert ^{2}\frac{\partial w}{\partial \overline{\phi }}=B^{2}f%
\frac{\partial f}{\partial \overline{\phi }}+\frac{\partial U}{\partial 
\overline{\phi }}\text{.}  \label{eq:Higgs}
\end{equation}%
Here, $B=\epsilon _{jk}\partial _{j}A_{k}$\ represents the magnetic field.

In order to obtain the first order self-dual equations of the model (\ref{1}%
), we start from the expression for the generalized total energy, i.e.,%
\begin{equation}
E=\int \left( \frac{1}{2}f^{2}B^{2}+w\left\vert D_{k}\phi \right\vert
^{2}+U\right) d^{2}x\text{,}
\end{equation}%
which can also be written in the form%
\begin{eqnarray}
E &=&\int \left( \frac{1}{2}\left( fB\mp \sqrt{2U}\right) ^{2}+w\left\vert
D_{\pm }\phi \right\vert ^{2}\pm B\left( f\sqrt{2U}\right) \right.   \notag
\\
&&\text{\ \ }\left. \frac{{}}{{}}\pm iw\epsilon _{ik}\left( \partial
_{i}\phi \right) \left( \partial _{k}\overline{\phi }\right) \mp ew\epsilon
_{ik}A_{k}\partial _{i}\left\vert \phi \right\vert ^{2}\right) d^{2}x\text{.}
\end{eqnarray}%
The energy is minimized by imposing%
\begin{equation}
D_{\pm }\phi =0\text{ \ \ and \ \ }B=\pm \frac{\sqrt{2U}}{f}\text{,}
\label{selfd1}
\end{equation}%
which are the generalized self-dual BPS equations. Considering Eqs. (\ref%
{selfd1}), the BPS energy is then reduced to%
\begin{eqnarray}
E_{BPS} &=&\pm \int \left( \frac{{}}{{}}\epsilon _{ik}\partial
_{i}A_{k}\left( f\sqrt{2U}\right) \right.   \label{xcxcx} \\
&&\left. \frac{{}}{{}}-ew\epsilon _{ik}A_{k}\partial _{i}\left\vert \phi
\right\vert ^{2}+iw\epsilon _{ik}\left( \partial _{i}\phi \right) \left(
\partial _{k}\overline{\phi }\right) \right) d^{2}x\text{,}  \notag
\end{eqnarray}%
and the static Amp\`{e}re's law is rewritten\ as%
\begin{equation}
\partial _{k}\left( f\sqrt{2U}\right) =-ew\partial _{k}\left\vert \phi
\right\vert ^{2}\text{.}  \label{cond_1}
\end{equation}%
With it,\ the BPS energy becomes%
\begin{equation}
E_{BPS}=\pm \int \left( \epsilon _{ik}\partial _{i}\left( A_{k}f\sqrt{2U}%
\right) +iw\epsilon _{ik}\left( \partial _{i}\phi \right) \left( \partial
_{k}\overline{\phi }\right) \right) d^{2}x\text{.}  \label{energy_BPS}
\end{equation}%
The point to be clarified here is that the integrand in (\ref{energy_BPS})
can be reduced to a total derivative only when considering axially symmetric
configurations. In this context, Eq. (\ref{cond_1}) stands for the key
condition for attaining self-duality in this generalized Maxwell-Higgs
theory.

Hence, from now on, the fields are supposed to be described by the usual
axially symmetric vortex Ansatz%
\begin{equation}
\phi \left( r,\theta \right) =vg\left( r\right) e^{in\theta }\text{ \ \ and
\ \ }\mathbf{A}\left( r,\theta \right) =-\frac{\widehat{\theta }}{er}\left(
a\left( r\right) -n\right) \text{,}  \label{2b}
\end{equation}%
where $n=\pm 1,\pm 2,\pm 3...$\ stands for the vorticity of the resulting
configuration, and the magnetic field is%
\begin{equation}
B\left( r\right) =-\frac{1}{er}\frac{da}{dr}\text{.}
\end{equation}%
The profile functions $g\left( r\right) $\ and $a\left( r\right) $\ are
constrained to behave according to the standard boundary conditions%
\begin{eqnarray}
g\left( 0\right)  &=&0\text{ \ \ and \ \ }g\left( \infty \right) =1\text{,}
\label{3} \\
a\left( 0\right)  &=&n\text{ \ \ and \ \ }a\left( \infty \right) =0\text{,}
\label{4}
\end{eqnarray}%
giving rise to regular solutions possessing finite energy, as desired.

Now, we come back to Eq. (\ref{energy_BPS}) defining it in terms of the
energy density $\varepsilon _{bps}$\ related to the BPS solutions as%
\begin{equation}
E_{bps}=\int \varepsilon _{bps}d^{2}x  \label{XXX}
\end{equation}%
where%
\begin{equation}
\varepsilon _{bps}=\mp \frac{1}{er}\frac{dH}{dr}\text{,}  \label{11}
\end{equation}%
with the auxiliary function $H\left( r\right) $\ being given by%
\begin{equation}
H\left( r\right) \equiv af\sqrt{2U}\text{.}  \label{12}
\end{equation}%
This function is finite at origin, $H\left( 0\right) =H_{0}$, and fulfills\ $%
H\left( \infty \right) =0$. \ Observing these boundary conditions for $%
H\left( r\right) ,$\ the resulting total energy (\ref{XXX})\ is%
\begin{equation}
E_{bps}=\frac{2\pi }{e}\left\vert H_{0}\right\vert \text{.}  \label{10}
\end{equation}%
One also remarks that $H_{0}$\ is proportional to $n$, the winding number
characterizing the vortex solution.

In terms of $g\left( r\right) $\ and $a\left( r\right) $, the BPS equations
read%
\begin{eqnarray}
\frac{dg}{dr} &=&\pm \frac{ag}{r}\text{,}  \label{8} \\
B &=&\pm \frac{\sqrt{2U}}{f}\text{.}  \label{9}
\end{eqnarray}%
which solve the Euler-Lagrange equations of motion. In order to perform such
verification\ explicitly, we first write the Amp\`{e}re's law (\ref%
{eq:Ampere}) in its axially symmetric form%
\begin{equation}
\frac{d}{dr}\left( f^{2}B\right) =-2ev^{2}w\frac{g^{2}a}{r}\text{,}
\label{ampere}
\end{equation}%
which becomes%
\begin{equation}
\frac{d}{dr}\left( f\sqrt{2U}\right) =-2ev^{2}wg\frac{dg}{dr}\text{,}
\label{cond_2}
\end{equation}%
when Eqs. (\ref{8}) and (\ref{9}) are used.\ The form (\ref{cond_2})
recovers the very same condition (\ref{cond_1})\ that assures the
self-duality of the overall model, revealing the consistence of the
self-dual equations with the Amp\`{e}re's law. In addition, one can express
Eq. (\ref{eq:Higgs})\ for the Higgs field in terms of $g\left( r\right) $\
and $a\left( r\right) $, that is,%
\begin{eqnarray}
&&\frac{d^{2}g}{dr^{2}}+\frac{1}{r}\frac{dg}{dr}-\frac{a^{2}g}{r^{2}}+\frac{1%
}{2w}\left( \left( \frac{dg}{dr}\right) ^{2}-\frac{a^{2}g^{2}}{r^{2}}\right) 
\frac{dw}{dg}  \notag \\
&=&\frac{1}{2wv^{2}}\left( B^{2}f\frac{df}{dg}+\frac{dU}{dg}\right) \text{.}
\label{higgs}
\end{eqnarray}%
It simply provides%
\begin{equation}
U=\frac{1}{2}f^{2}B^{2}\text{,}
\end{equation}%
when saturated by the self-dual equations, which coincides with Eq. (\ref{9}%
). In this way, we have\ explicitly shown that the self-dual equations solve
the stationary Euler-Lagrange equations of motion.

Due to the arbitrariness of $f\left( g\right) $\ and $w\left( g\right) $,
the search for solutions to the axially symmetric Euler-Lagrange equations
can be a quite hard task, even in the presence of the suitable boundary
conditions (\ref{3}) and (\ref{4}). A way to circumvent this point is
focusing the attention on the self-dual equations (\ref{8}) and (\ref{9}).
However, it is worthwhile to reinforce that such equations only hold when
the model is constrained by condition (\ref{cond_1}), also expressed as%
\begin{equation}
\frac{d}{dg}\left( f\sqrt{2U}\right) =-2ev^{2}wg\text{.}  \label{6}
\end{equation}%
We can summarize in the following way: given a set of functions $f$, $w$\
and $U$\ satisfying (\ref{6}), regular solutions $g\left( r\right) $\ and $%
a\left( r\right) $\ can be found by solving (\ref{8}) and (\ref{9}) using
the boundary conditions (\ref{3}) and (\ref{4}). The resulting
configurations stand for topological vortices possessing finite energy given
by\ (\ref{10}), which remains proportional to the magnetic flux $\Phi
_{B}=2\pi n/e$. The proportionality constant is finite and related to the
value of $f\sqrt{2U}$\ near the origin. Moreover, it is worthwhile to point
out that 
\begin{equation}
\varepsilon _{bps}=2U+2v^{2}w\left( \frac{ag}{r}\right) ^{2}\text{,}
\end{equation}%
is the BPS energy density (\ref{11}), which becomes positive whenever Eq. (%
\ref{6}) ensures\ a positive $w$ (for a given pair of functions $U$\ and $f$%
\ conveniently chosen).

The next Section introduces some effective Maxwell-Higgs models for which
the BPS equations (\ref{8}) and (\ref{9}) can be solved analytically
(instead of numerically, as usually done). The analytical profiles
representing $g\left( r\right) $, $a\left( r\right) $, $B\left( r\right) $
and $\varepsilon _{bps}$ (\ref{11}) are depicted and compared with the usual
(numerical) ANO\ solution. Furthermore, the main features of the new
analytical vortices are highlighted.


\section{Analytical BPS vortices}

Now, we present the main goal of this work by introducing generalized
Maxwell-Higgs models for which the BPS equations (\ref{8}) and (\ref{9}) can
be solved analytically according the finite energy boundary conditions (\ref%
{3}) and (\ref{4}). Here, for simplicity, we only consider those
configurations for which the winding number\ is equal to the unity $\left(
n=1\right) $, although it is also possible to find solutions with higher
vorticity, as it will be explained below.

Along this section, we work with the upper signs in Eqs. (\ref{8}), (\ref{9}%
) and (\ref{11}) only. Also, for simplicity, we set $e=v=1$. Our
prescription to find analytical self-dual vortices can be described as
follows. Firstly, we choose the potential $U\left( g\right) $\ supporting
the spontaneous symmetry breaking of the $U(1)$ local gauge symmetry
inherent to the model (\ref{1}). In the sequel, we choose an analytical
function $g\left( r\right) $\ satisfying the boundary conditions (\ref{3}).
Then, we use $g\left( r\right) $\ to solve Eq. (\ref{8}), which allows to
obtain the corresponding profile for $a\left( r\right) $\ fulfilling the
boundary conditions (\ref{4}). In the end, we use Eq. (\ref{9}) to evaluate
the expression for the generalizing function $f$, writing it as a function
of the radial variable $r$\ (i.e., regarding the analytical models as
effective ones).

A general observation about the kinetic functions $f^{2}\left( \left\vert
\phi \right\vert \right) $\ and $w\left( \left\vert \phi \right\vert \right) 
$\ is that they are presented as functions of the radial variable $r$, not
of the field variable $g$, i.e., $|\phi |$. Expressing $f$\ and $w$\ in
terms of $g$\ gives very long expressions, when possible. In general, it
becomes a very difficult task.

The analytical profiles here obtained provide a set of self-dual vortices
possessing finite total energy given by Eq. (\ref{10}). It is also
worthwhile to remember that the corresponding $f$'s and $w$'s are positive,
as required. These new solutions are shown in figs. 1, 2, 3 and 4, from
which we highlight their main features.

\subsection{$\left\vert \protect\phi \right\vert ^{4}$-models}

We first investigate some models defined by the usual fourth-order Higgs
potential%
\begin{equation}
U\left( g\right) =\frac{1}{2}\left( 1-g^{2}\right) ^{2}\text{,}  \label{13}
\end{equation}%
where the constant the scalar-matter self-interaction was supposed equal to
the unity, for simplicity.

The $\left\vert \phi \right\vert ^{4}$-models here presented possess a
generalizing function $f\left( r\right) $\ finite at the boundaries (i.e.,
for $r=0$\ and asymptotically). In this case, the BPS equations (\ref{8})
and (\ref{9}) reduce to%
\begin{eqnarray}
\frac{dg}{dr} &=&\frac{ag}{r}\text{,}  \label{13.1} \\
\frac{1}{r}\frac{da}{dr} &=&\frac{g^{2}-1}{f}\text{.}  \label{13.2}
\end{eqnarray}%
In particular, one clearly sees that $f=1$ leads us back to the standard case%
\begin{eqnarray}
\frac{dg}{dr} &=&\frac{ag}{r}\text{,} \\
\frac{1}{r}\frac{da}{dr} &=&g^{2}-1\text{,}
\end{eqnarray}%
which yields the well-known Abrikosov-Nielsen-Olesen numerical vortices \cite%
{n1}. Moreover, families containing nontrivial numerical solutions of the
same kind were also studied in Ref. \cite{gv}.

We now proceed looking for analytical solutions of\ Eqs. (\ref{13.1}) and (%
\ref{13.2}). In this sense, by following our prescription, the first model
we introduce is defined by the BPS\ solution%
\begin{equation}
g\left( r\right) =\tanh \left( r\right) \text{,}  \label{1g}
\end{equation}%
which trivially obeys (\ref{3}). Replacing it in Eq. (\ref{13.1}), one
achieves the gauge field profile%
\begin{equation}
a\left( r\right) =\frac{2r}{\sinh \left( 2r\right) }\text{,}  \label{1a}
\end{equation}%
satisfying the boundary conditions (\ref{4}), with $n=1$. Now, by using\
Eqs. (\ref{1g}) and (\ref{1a}) in Eq. (\ref{13.2}), one gets that the
corresponding function $f\left( r\right) $ reads as%
\begin{equation}
f\left( r\right) =\frac{r\left( 1-\cosh \left( 2r\right) \right) }{\sinh
\left( 2r\right) -2r\cosh \left( 2r\right) }\text{,}  \label{1f}
\end{equation}%
being a smooth and positive function with values $f\left( 0\right) =3/4$\
and $f\left( \infty \right) =1/2$. Finally, we obtain the auxiliary function 
$H$\ by substituting Eqs. (\ref{13}), (\ref{1a}) and (\ref{1f}) in Eq. (\ref%
{12}). The resulting expression is%
\begin{equation}
H\left( r\right) =\frac{r^{2}\sinh \left( r\right) }{\left( \frac{1}{2}\sinh
\left( 2r\right) -2r\cosh ^{2}\left( r\right) +r\right) \cosh ^{3}\left(
r\right) }\text{,}  \label{1h}
\end{equation}%
a smooth function for which the values at the boundaries read as $H\left(
0\right) =-3/4$\ and$\ H\left( \infty \right) =0$, as desired. Finally, the
magnetic field associated with (\ref{1a})\ is given by%
\begin{equation}
B\left( r\right) =\frac{2\left( 2r\cosh \left( 2r\right) -\sinh \left(
2r\right) \right) }{r\sinh ^{2}\left( 2r\right) }\text{,}  \label{1b}
\end{equation}%
whose profile is a lump centered at the origin; see the dash-dotted red line
in Fig. 3. Therefore,\ the solutions (\ref{1g}) and (\ref{1a}) represent
analytical BPS Maxwell-Higgs vortices possessing total energy equal to $%
E_{bps}=3\pi /2$, according to the Eq. (\ref{10}).

The second model, inherent to the $\left\vert \phi \right\vert ^{4}$%
-potential, is defined by the BPS\ solution%
\begin{equation}
g\left( r\right) =\sqrt{1-e^{-r^{2}}}\text{,}  \label{2g}
\end{equation}%
which also fulfills the conditions (\ref{3}). From Eq. (\ref{13.1}), we
attain\ the gauge field profile%
\begin{equation}
a\left( r\right) =\frac{r^{2}}{e^{r^{2}}-1}\text{,}  \label{02a}
\end{equation}%
satisfying the boundary conditions (\ref{4}) (also with $n=1$). Then, by
combining\ Eqs. (\ref{2g}) and (\ref{02a})\ in Eq. (\ref{13.2}),\ the
following generalizing function is achieved:%
\begin{equation}
f\left( r\right) =\frac{\left( e^{-r^{2}}-1\right) ^{2}}{2\left(
e^{-r^{2}}+r^{2}-1\right) }\text{.}  \label{2f}
\end{equation}%
This is a positive and finite function whose values at the boundaries are $%
f\left( r=0\right) =1$\ and $f\left( r=\infty \right) =0$. Moreover, the
corresponding $H\left( r\right) $\ is given by%
\begin{equation}
H\left( r\right) =\frac{r^{2}e^{-2r^{2}}\left( e^{-r^{2}}-1\right) }{2\left(
e^{-r^{2}}+r^{2}-1\right) }\text{,}  \label{2h}
\end{equation}%
providing $H\left( 0\right) =-1$\ and $H\left( \infty \right) =0$, as
required. The magnetic field,%
\begin{equation}
B\left( r\right) =\frac{2e^{-r^{2}}\left( e^{-r^{2}}+r^{2}-1\right) }{\left(
e^{-r^{2}}-1\right) ^{2}}\text{,}  \label{02b}
\end{equation}%
also is a lump centered at the origin; see the dashed green line in Fig. 3.
This way, the profiles Eqs. (\ref{2g}) and (\ref{02a}) describe analytical
self-dual vortices whose BPS total energy is $E_{bps}=2\pi $. In particular,
it means that these analytical solutions saturate the very same Bogomol'nyi
bound fulfilled by the usual (numerical) $n=1$\ ANO vortex.

\subsection{$\left\vert \protect\phi \right\vert ^{6}$-models}

As it was shown in Ref. \cite{gv}, the generalized model allows to find BPS
vortices even in the presence of a higher order potential describing the
scalar-matter self-interaction. In this sense, we now go further by
introducing analytical self-dual vortices arising in the presence of a
sixth-order potential, defined by%
\begin{equation}
U\left( g\right) =\frac{1}{2}g^{2}\left( 1-g^{2}\right) ^{2}\text{.}
\label{6p}
\end{equation}%
The vacuum manifold of the corresponding $\left\vert \phi \right\vert ^{6}$%
-model is represented by a dot surrounded by a circle, the dot standing for
a symmetric vacuum. As a consequence, under a suitable choose of the
boundary conditions to be satisfied by the profile functions $g\left(
r\right) $\ and $a\left( r\right) $, the model also supports nontopological
self-dual structures possessing finite energy. \ Indeed, some of us have
already obtained such objects, with results being forthcoming reported \cite%
{ntv}. It is worthwhile to point out that nontopological vortices do not
occur in the $\left\vert \phi \right\vert ^{4}$-model (\ref{13}) because it
has no a symmetric vacuum (i.e., its vacuum manifold is a circle).

On the other hand, it is well-known that the $\left\vert \phi \right\vert
^{6}$-potential (\ref{6p}) ensures the self-duality of the usual
Chern-Simons-Higgs (CSH) model, whose topological vortices possess both
electric and magnetic fields. Despite our generalized $\left\vert \phi
\right\vert ^{6}$-Maxwell-Higgs model supports only noncharged self-dual
solutions, they are expected to behave in the same general way the CSH ones
do. Some numerical self-dual $\left\vert \phi \right\vert ^{6}$%
-Maxwell-Higgs models were already investigated in Ref. \cite{gv}. Here, for
completeness, we consider only the analytical solutions for it.

Returning to our prescription, under the $\left\vert \phi \right\vert ^{6}$%
-potential (\ref{6p}), the BPS equations (\ref{8}) and (\ref{9}) can be
written as%
\begin{eqnarray}
\frac{dg}{dr} &=&\frac{ag}{r}\text{,}  \label{bps3} \\
\frac{1}{r}\frac{da}{dr} &=&\frac{g\left( g^{2}-1\right) }{f}\text{.}
\label{bps4}
\end{eqnarray}%
In addition, as we demonstrate below, the resulting generalizing function $%
f\left( r\right) $\ is not necessarily finite at the boundaries.

The first analytical model\ has the profile $g\left( r\right) $\ defined by%
\begin{equation}
g\left( r\right) =\frac{r}{\sqrt[4]{1+r^{4}}}\text{,}  \label{g16}
\end{equation}%
from which one gets, according to Eq. (\ref{bps3}), the corresponding gauge
solution,%
\begin{equation}
a\left( r\right) =\frac{1}{1+r^{4}}\text{,}  \label{a16}
\end{equation}%
whilst Eq. (\ref{bps4}) gives the following generalizing function%
\begin{equation}
f\left( r\right) =\frac{\left( 1+r^{4}\right) ^{\frac{5}{4}}\left( \sqrt{%
1+r^{4}}-r^{2}\right) }{4r}\text{.}  \label{f16}
\end{equation}%
It has the following behavior at the boundaries: $f\left( 0\right) =f\left(
\infty \right) =\infty $. However, even in this case, one still achieves
well-behaved solutions possessing finite total energy. To clarify the way it
happens, we calculate the fields $B\left( r\right) $\ and $H\left( r\right) $%
\ arising from\ Eqs. (\ref{a16}) and (\ref{g16}). The magnetic field,%
\begin{equation}
B\left( r\right) =\frac{4r^{2}}{\left( 1+r^{4}\right) ^{2}}\text{,}
\label{b16}
\end{equation}%
\ presents a ringlike profile, a typical magnetic behavior related to the $%
\left\vert \phi \right\vert ^{6}$-vortices; see the dotted blue line in Fig.
3. The auxiliary function $H\left( r\right) $\ reads as%
\begin{equation}
H\left( r\right) =-\frac{\left( \sqrt{1+r^{4}}-r^{2}\right) ^{2}}{4\sqrt{%
1+r^{4}}}\text{,}  \label{h16}
\end{equation}%
which, at the boundaries, assumes the values\ $H\left( 0\right) =-1/4$\ and $%
H\left( \infty \right) =0$, as desired. As a result, the analytical profiles
(\ref{g16}) and (\ref{a16}) give rise to a BPS\ vortex whose total energy is 
$E_{bps}=\pi /2$.

The second $\left\vert \phi \right\vert ^{6}$-model\ is a little bit more
sophisticated than the previous ones, the Higgs profile\ being given by%
\begin{equation}
g\left( r\right) =e^{-\frac{1}{4}\mbox{Ei}\left( 1,\frac{1}{4}r^{4}\right) }%
\text{,}  \label{ug}
\end{equation}%
with the function $\mbox{Ei}\left( 1,r\right) $\ standing for the
exponential integral%
\begin{equation}
\mbox{Ei}\left( 1,r\right) \equiv \overset{\infty }{\underset{1}{\int }}%
\frac{e^{-rx}}{x}dx\text{.}
\end{equation}%
Nevertheless, the gauge field has a simpler solution,%
\begin{equation}
a\left( r\right) =e^{-\frac{1}{4}r^{4}}\text{,}
\end{equation}%
whereas the corresponding functions\ $f\left( r\right) $\ and $H\left(
r\right) $ are defined by%
\begin{equation}
f\left( r\right) =-\frac{e^{-\frac{1}{2}\mbox{Ei}\left( 1,\frac{1}{4}%
r^{4}\right) }-1}{r^{2}}e^{-\frac{1}{4}\mbox{Ei}\left( 1,\frac{1}{4}%
r^{4}\right) +\frac{1}{4}r^{4}}\text{,}
\end{equation}%
\begin{equation}
H\left( r\right) =-\frac{\left( e^{-\frac{1}{2}\mbox{Ei}\left( 1,\frac{1}{4}%
r^{4}\right) }-1\right) ^{2}}{r^{2}}e^{-\frac{1}{2}\mbox{Ei}\left( 1,\frac{1%
}{4}r^{4}\right) }\text{,}
\end{equation}%
from which we get that $f\left( 0\right) =\infty $\ and $f\left( \infty
\right) =0$, whilst $H\left( \infty \right) =0$\ and $H\left( 0\right) =-%
\frac{1}{2}\sqrt{e^{\gamma }}$, $\gamma $\ being the Euler's constant ($%
\gamma =0.5772156649...$). We see that $f\left( r\right) $\ is divergent at
the origin, whilst vanishing asymptotically. Notwithstanding, as in the
previous case, the Bogomol'nyi bound for the BPS\ total energy saturates at $%
E_{bps}=\pi \sqrt{e^{\gamma }}$. The magnetic field of the resulting
self-dual configuration,%
\begin{equation}
B\left( r\right) =r^{2}e^{-\frac{1}{4}r^{4}}\text{,}
\end{equation}%
also presents the aforecited typical ringlike behavior; see the long-dashed
gold line in Fig. 3.

In the sequel, we depict all the analytical solutions together with the
standard ANO profile, from which we highlight their main features and also
the differences of the generalized solutions in comparison with the usual
Maxwell-Higgs ones.

\begin{figure}[]
\begin{center}
\scalebox{1}[1]{\includegraphics[width=8.5cm]{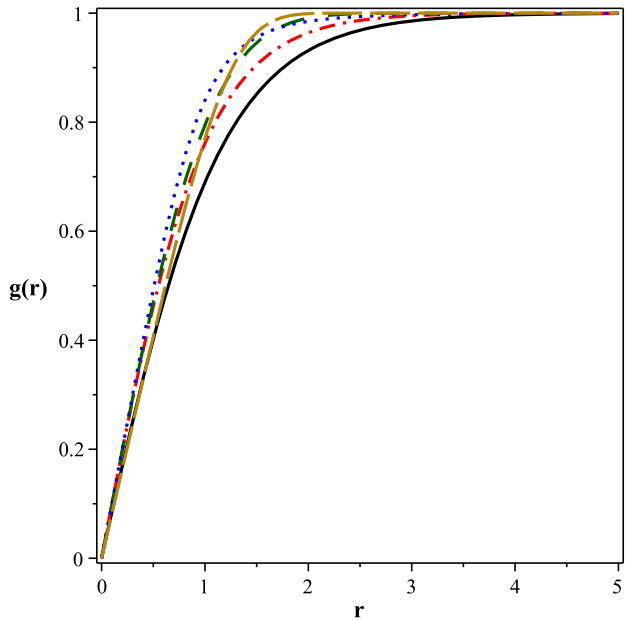}}
\end{center}
\par
\vspace{-0.5cm}
\caption{Solutions to $g\left( r\right) $ given by (\protect\ref{1g}){\
(dash-dotted red line), (\protect\ref{2g}) (dashed green line), (\protect\ref%
{g16}) (dotted blue line) and (\protect\ref{ug}) (long-dashed gold line).
Here, the solid black line is the }standard (numerical) $n=1$ ANO solution.}
\end{figure}

The analytical solutions defining the Higgs profiles $g\left( r\right) $\
are depicted in Fig. 1. The profile (\ref{1g}) is plotted with the
dash-dotted red line, whilst Eq. (\ref{2g})\ is represented by the dashed
green line; these solutions corresponding to the generalized $\left\vert
\phi \right\vert ^{4}$-models. \ On the other hand, the dotted blue line
stands for (\ref{g16}) and the long-dashed gold line represents (\ref{ug});
both belonging to the noncanonical $\left\vert \phi \right\vert ^{6}$%
-models. The standard (numerical) $n=1$\ ANO Higgs profile is drawn with the
solid black line. The overall conclusion is that the analytical solutions
behave in the same general way the standard one does. However, the new
profiles saturate the asymptotic value $g\left( r=\infty \right) =1$\
faster, so that the new Higgs profiles are more localized, and the
corresponding bosons are more massive than the ANO ones. \ From now on, we
follow\ the same line/color definitions established in Fig. 1.

\begin{figure}[]
\begin{center}
\scalebox{1}[1]{\includegraphics[width=8.5cm]{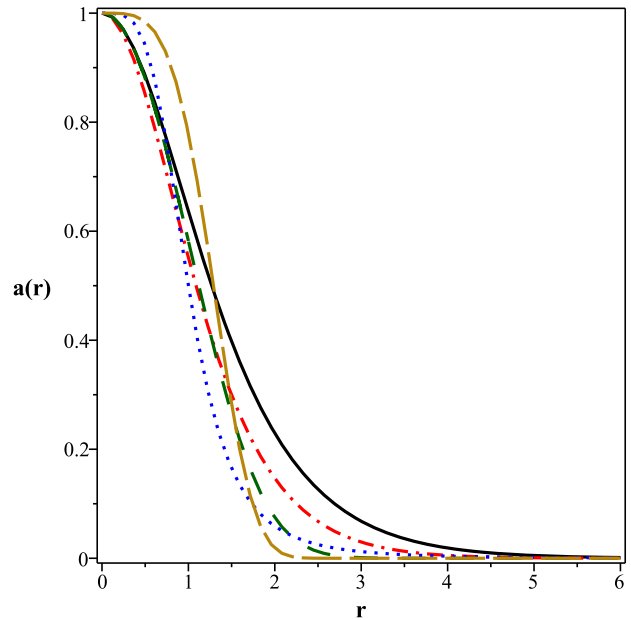}}
\end{center}
\par
\vspace{-0.5cm}
\caption{Solutions to $a\left( r\right) $. Conventions as in Fig. 1.}
\end{figure}

The gauge profiles\ $a\left( r\right) $\ are plotted in Fig. 2. There, we
see that the profiles (\ref{g16}) and (\ref{ug}), related to the $\left\vert
\phi \right\vert ^{6}$-potential, have developed a plateau close to the
origin, such structure being a common feature presented in the self-dual $%
\left\vert \phi \right\vert ^{6}$-scenarios. On the other hand, the profiles
(\ref{1a}) and (\ref{02a}), corresponding to the $\left\vert \phi
\right\vert ^{4}$-models, are lumps centered at the origin. The amplitude in 
$r=0$\ corresponds to the winding number $n=1$, as\ already commented. For
large radius, all profiles vanish monotonically.

We show the profiles we have found for the magnetic field $B\left( r\right) $
in Fig. 3. The solutions regarding the fourth-order potential are lumps
centered at the origin, just as the ANO magnetic field. On the other hand,
the solutions related to the sixth-order potential present a ringlike
behavior, as expected when considering the magnetic fields belonging to the
self-dual $\left\vert \phi \right\vert ^{6}$-vortices. We also observe
differences both on the amplitudes and on the characteristic lengths of the
magnetic profiles.

\begin{figure}[]
\begin{center}
\scalebox{1}[1]{\includegraphics[width=8.5cm]{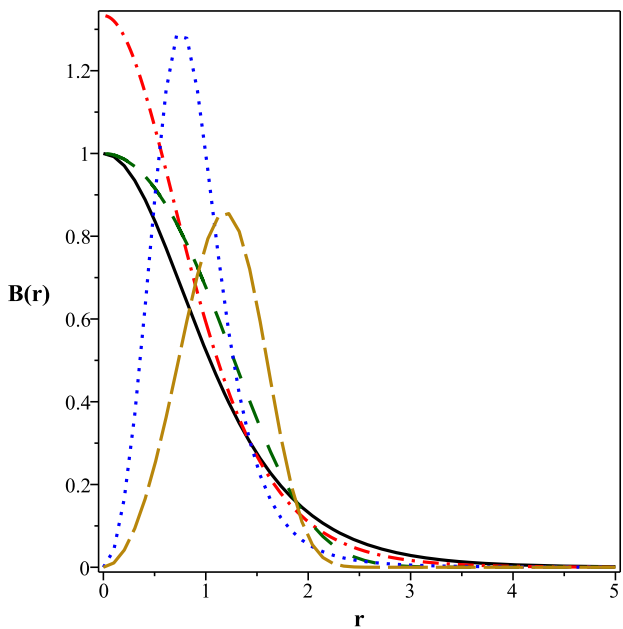}}
\end{center}
\par
\vspace{-0.5cm}
\caption{Solutions to $B\left( r\right) $. Conventions as in Fig. 1.}
\end{figure}

\begin{figure}[]
\begin{center}
\scalebox{1}[1]{\includegraphics[width=8.5cm]{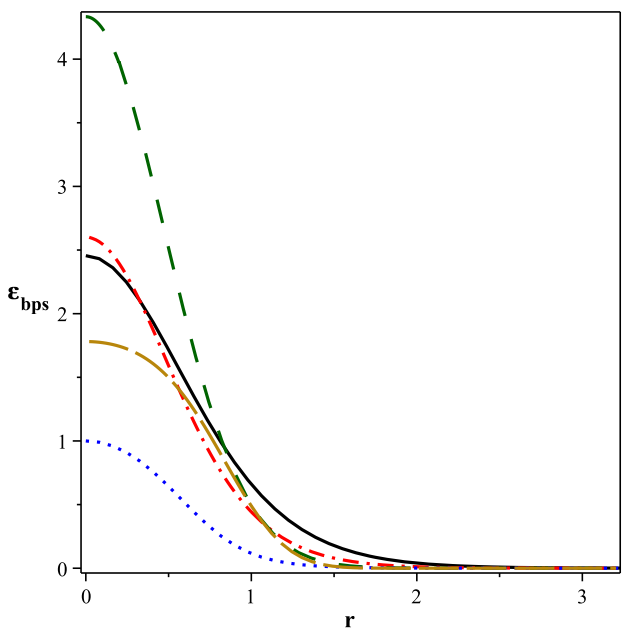}}
\end{center}
\par
\vspace{-0.5cm}
\caption{Solutions to $\protect\varepsilon _{bps}$. Conventions as in Fig.
1. }
\end{figure}

By last, the BPS energy densities are depicted in Fig. 4. We see that all
the profiles are lumps centered at origin, as expected for $n=1$ vortices
(including the $\left\vert \phi \right\vert ^{6}$-ones). Besides the
different characteristic lengths, it is also interesting to note that the
profiles coming from the fourth-order potential have achieved greater
amplitudes than those ones related to the sixth-order potential.

In the next Section, we present our final considerations and the
perspectives regarding future works.


\section{Ending comments}

\label{end}

We have investigated the existence of analytical BPS vortices within\ the
nonstandard Maxwell-Higgs scenario proposed in Ref. \cite{gv}, such model
being generalized by two positive functions, namely $f\left( \left\vert \phi
\right\vert \right) $ and $w\left( \left\vert \phi \right\vert \right) $,
which change the overall dynamics of the original theory; see Eq.(\ref{1}).
By imposing the radially symmetric ansatz (\ref{2b}) for the gauge and
scalar fields, we have reviewed the BPS framework and achieved the
first-order equations whose corresponding solutions have finite energy. The
self-duality arises when the condition (\ref{6}) is satisfied by the
potential $U$ and the functions $f$ \ and $w$.

The existence of a well-defined Bogomol'nyi bound (\ref{10}) is ruled by an
auxiliary function $H\left( r\right) $ obeying appropriated boundary
conditions which guarantee a finite total energy. So, the well-behaved $%
H\left( r\right) $ selects the function $f\left( r\right) $ defining the
generalized model. With this prescription, we have obtained\ analytical
profiles to some $n=1$ self-dual vortex configurations\ within both $%
\left\vert \phi \right\vert ^{4}$- and $\left\vert \phi \right\vert ^{6}$%
-models saturating different energy bounds.

In order to compare the analytical self-dual solutions with the ANO ones, we
have depicted the profiles for the scalar, gauge and magnetic fields,\ in
figs. 1, 2 and 3, respectively, whilst the BPS energy density is plotted in
Fig. 4. The overall conclusion is that the new profiles are well-behaved in
all relevant sectors, assuring the consistence of the generalized models
here proposed. Furthermore, all the solutions have provided localized energy
densities and magnetic fields, as expected. In general, they mimic the
behavior of the well-known numerical\ configurations,\ as\ the ringlike
magnetic field related to the self-dual $\left\vert \phi \right\vert ^{6}$%
-vortices.

Concerning the possibility of obtaining analytical solutions with higher
winding numbers, no obvious route for it seems to be available within the
models we have studied in this work. Nevertheless, it is possible to
construct analytical BPS vortices possessing higher vorticity but different
generalizing functions $f\left( r\right) $. Such procedure is clarified by
the following example related to the $\left\vert \phi \right\vert ^{6}$%
-models. We propose%
\begin{equation}
g\left( r\right) =\frac{r^{n}}{\left( 1+r^{m}\right) ^{n/m}}\text{,}
\end{equation}%
with $m$, $n>0$, as a generalization for the Higgs profile given by Eq. (\ref%
{g16}).\ This leads to%
\begin{equation}
a(r)=\frac{n}{1+r^{m}}\text{,}
\end{equation}%
compatible with the boundary conditions for a vortex possessing any integer
winding number (i.e., $n=+1,+2,+3...$). The resulting magnetic field is%
\begin{equation}
B(r)=\frac{mnr^{m-2}}{\left( 1+r^{m}\right) ^{2}}\text{,}
\end{equation}%
whilst the corresponding $f\left( r\right) $\ and $H\left( r\right) $\ are%
\begin{equation}
f\left( r\right) =\frac{r^{n-m+2}}{nm}R^{2-3n/m}\left(
R^{2n/m}-r^{2n}\right) ^{2}\text{,}  \label{xx}
\end{equation}%
\begin{equation}
H\left( r\right) =\frac{r^{2n-m+2}}{m}R^{1-6n/m}\left(
R^{2n/m}-r^{2n}\right) ^{2}\text{,}
\end{equation}%
respectively, with $R=1+r^{m}$. However, only the case $m=2n+2$\ provides
positive $f$\ and $w$, nonsingular magnetic field and finite energy.
Obviously, the choice $n=1$ and $m=4$\ reduces this general proposal to the
case (\ref{g16}).\ The generalization of the vortex configurations
presenting higher winding numbers within other field scenarios can follow
this general idea.

Regarding future works, interesting issues include the search for the
nontopological self-dual vortices arising in the generalized Maxwell-Higgs
scenario (\ref{1}) when endowed by the sixth-order potential (\ref{6p}); see
Ref. \cite{ntv}. In parallel, some of us are also working in a general
formulation of the deformation method \cite{ll} applicable to field models
possessing generalized dynamics \cite{df}. These two fronts are now under
investigation, and we expect interesting results for a future report.

The Brazilian authors thank CAPES, CNPq and FAPEMA for partial financial
support.

\end{document}